\newcommand{\mr}[1]{\ensuremath{\mathrm{#1}}}

\documentclass[aps,prb,reprint,amsmath,amssymb,superscriptaddress]{revtex4-1}
\usepackage{graphicx}	 % Use pdf, png, jpg, or eps\UTF{00DF} with pdflatex; use eps in DVI mode

\usepackage{color}
\usepackage{dcolumn}% Align table columns on decimal point
\usepackage{bm}% bold math
\usepackage[mathlines]{lineno}% Enable numbering of text and display math
\begin{document}

%\preprint{}

\title{Picosecond strain dynamics in $\mathbf{Ge_{2}Sb_{2}Te_{5}}$ monitored by time-resolved x-ray diffraction}


\author{Paul Fons}
%\thanks{}
\email[]{paul-fons@aist.go.jp}
\affiliation{Nanoelectronics Research Institute, AIST, Tsukuba Central 4, Higashi 1-1-1,Tsukuba, Ibaraki, 305-8562, Japan}
\affiliation{SPring8, Japan Synchrotron Radiation Institute (JASRI), Kouto 1-1-1, Sayo-cho, Sayo-gun, Hyogo 679-5148, Japan}
\author{Peter Rodenbach}
\affiliation{Paul-Drude-Institut f\"{u}r Festk\"{o}rperelektronik, Hausvogteiplatz 5-7, 10117, Berlin, Germany}
\author{Kirill V. Mitrofanov}
\affiliation{Nanoelectronics Research Institute, AIST, Tsukuba Central 4, Higashi 1-1-1,Tsukuba, Ibaraki, 305-8562, Japan}
\author{Alexander V. Kolobov}
\affiliation{Nanoelectronics Research Institute, AIST, Tsukuba Central 4, Higashi 1-1-1,Tsukuba, Ibaraki, 305-8562, Japan}
\affiliation{SPring8, Japan Synchrotron Radiation Institute (JASRI), Kouto 1-1-1, Sayo-cho, Sayo-gun, Hyogo 679-5148, Japan}
\author{Junji Tominaga}
\affiliation{Nanoelectronics Research Institute, AIST, Tsukuba Central 4, Higashi 1-1-1,Tsukuba, Ibaraki, 305-8562, Japan}
\author{Roman Shayduk}
\affiliation{Helmholtz-Zentrum Berlin f\"{u}r Materialien und Energie GmbH,Wilhelm-Conrad-R\"{o}ntgen Campus, BESSY II ,Albert-Einstein-Str.\ 15, 12489 Berlin, Germany}
\affiliation{Deutsches Elektronen-Synchrotron,\\
Notkestrasse 85, 22607 Hamburg, Germany}
\author{Alessandro Giussani}
\author{Raffaella Calarco}
\author{Michael Hanke}
\author{Henning Riechert}
\affiliation{Paul-Drude-Institut f\"{u}r Festk\"{o}rperelektronik, Hausvogteiplatz 5-7, 10117, Berlin, Germany}
\author{Robert E. Simpson}
\affiliation{Singapore University of Technology and Design, Engineering Product Development, 138682 Singapore, Singapore}
\author{Muneaki Hase}
\affiliation{Institute of Applied Physics, University of Tsukuba, 1-1-1 Tennodai, Tsukuba 305-8573, Japan}

\date{\today}

\begin{abstract}
Coherent phonons (CP) generated by laser pulses on the femtosecond scale have been proposed as a means to achieve ultrafast, non-thermal switching in phase-change materials such as $\mathrm{Ge_{2}Sb_{2}Te_{5}}$~(GST).  Here we use ultrafast optical pump pulses to induce coherent acoustic phonons  and stroboscopically measure the corresponding lattice distortions in GST using 100 ps x-ray pulses from the ESRF storage ring.  A linear-chain model provides a good description of the observed changes in the diffraction signal, however, the magnitudes of the measured shifts are too large to be explained by thermal effects alone implying the presence of transient non-equilibrium electron heating in addition to temperature driven expansion. The information on the movement of atoms during the excitation process can lead to greater insight into the possibilities of using CP-induced phase-transitions in GST.
\end{abstract}

%\keywords{}

\maketitle

\section{Introduction}
Materials such as the prototypical phase-change compound $\mathrm{Ge_{2}Sb_{2}Te_{5}}$ (GST) are in wide spread use in optical storage as well as in electrical memory, a technology envisioned by many as a successor to FLASH memory.  In both cases, the large property contrast between the amorphous and crystalline phases is used to store information; these two states are also referred to as the high-resistivity RESET state and the low-resistivity SET state, respectively in electronic memory terminology. Detailed structural studies have demonstrated that the large electrical and optical contrast between the phases is a result of a correspondingly large difference between the nature of the bonding of the two structures which requires only local atomic rearrangements to occur. \cite{Shportko08,Kolobov04} In devices, switching between the SET and RESET state is typically achieved by applying intense, short, nanosecond-order electrical or optical pulses, while reverting from the RESET to the SET states is accomplished by applying longer, less intense pulses. Conventionally the SET state has been associated with the crystalline state while the RESET state has been associated with the amorphous state although recently a new variation of electrical phase change memory called interfacial phase-change memory (iPCM) has been demonstrated in which both the SET and RESET states are crystalline.\cite{Simpson11}  Although it is well known that the application of pulses of tens to hundreds of nanoseconds in duration leads to the formation of a glassy state generated via a quench from the molten state, the application of shorter duration pulses has been suggested to lead to a different amorphization process. Optical pump, x-ray probe experiments using sub-nanosecond pulses have suggested the presence of a transient state originating from intense optical excitation that is different than the equilibrium molten state. \cite{Fons10}  These reports suggest that appropriate local distortions induced in the presence of a strong optical pulse may lead to faster and possibly more efficient device switching.

Lattice vibrations (phonons) play an important part in many condensed matter processes including phase-transitions, bond hardening and softening, and melting to name a few.  Based upon an atomistic understanding of the switching process  in GST, it has been suggested that the selective generation of a transient state with the desired local distortions generated via excitation of a particular phonon mode(s) can provide fast and highly efficient pathways to selectively drive phase transitions eliminating the entropy related time and energy losses associated with switching via the melt-quench/recrystallization process.\cite{Kolobov11} Recently, selective generation of coherent phonons using a double-pump femtosecond laser was used to switch iPCM material from the RESET to the SET state at room-temperature, a process which in equilibrium typically requires heating the same material to over \mr{150^{\circ} C }.\cite{Makino11} By selective excitation of the $\mathrm{A_{1}}$ phonon mode, an iPCM sample was permanently transformed into the SET state with an average temperature rise of less than \mr{10^{\circ} \, C} (fluence \mr{0.1 \, mJ/cm^{2}}).\cite{Makino12}   In addition, femtosecond optical switching experiments on epitaxial films have also suggested alternative athermal crystallization processes may be present when ultrafast optical irradiation is used. \cite{RodenbachAPL}  These observations serve as motivation for the direct study of atom displacement in ultrafast laser-induced coherent phonon excitation.

\section{Time-resolved x-ray diffraction study of Coherent Acoustical Phonons}
\subsection{Experimental Setup}
\begin{figure}[h] %  figure placement: here, top, bottom, or page
   \centering
   \includegraphics[width=8.3 cm]{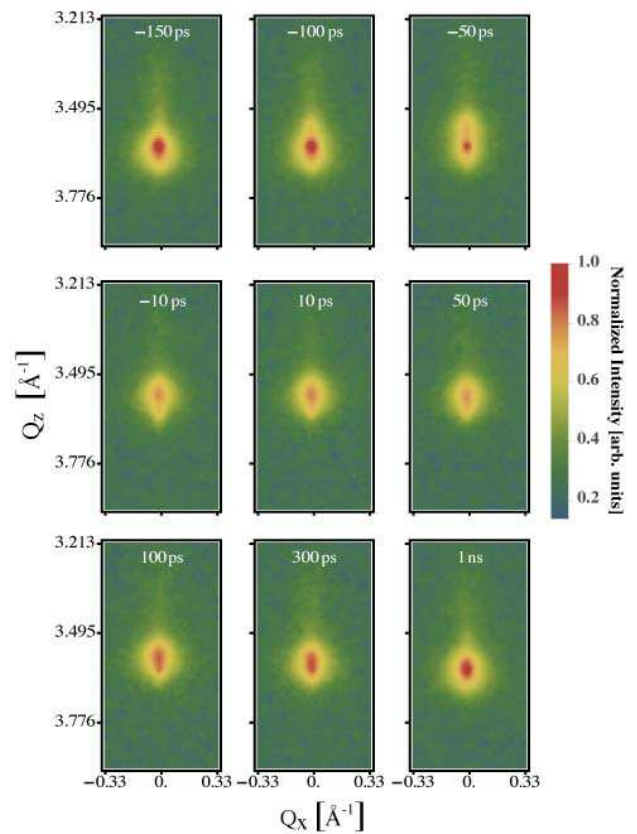} 
   \caption{(Color Online) A series of snapshots showing the time evolution of the GST (222) diffraction peak.  From the upper left hand corner across each row, the times are -150 ps, -100 ps, -50 ps, -10 ps, 10 ps, 50 ps, 100 ps, 300 ps, and 1 ns, respectively. The \mr{Q_{z}} axis in the figure corresponds to the reciprocal space direction $\left <111 \right >^{*}$ while the \mr{Q_{x}} axis corresponds to the reciprocal space $\left <110 \right >^{*}$ direction. The laser fluence used was {$\mathrm{16 \, mJ/cm^{2}}$}.}
\label{fig1}
\end{figure}

As a precursor to the control of coherent phonons in GST, we have investigated the generation of coherent acoustic phonons using a fast optical pump and synchrotron x-ray probe pulses. To eliminate grain boundaries, epitaxial GST samples grown on Si (111) were used; further details on the growth and characterization of the epilayers can be found in the literature. \cite{Rodenbach12} The film/substrate crystallographic relationship was $\mathrm{[111]_{F} \, \| \,[111]_{S} \, and \, [110]_{F} \, \| \, [110]_{S}}$.  In the current experiment, a 30 nm thick epitaxial GST film was irradiated in a \mr{80 \, \mu m} spot by a 700 fs, 800 nm laser pulse from a Ti-sapphire amplified laser system at ID9 at the ESRF.   Herein, we use the rocksalt structure for the Miller indices in all figures and text.
 Spatial overlap was ensured by alignment of both pump and probe pulses using a photodiode. T-bone by a 5 nm thick amorphous \mr{Si_{3}N_{4}} cap to prevent oxidation.  A 18 keV x-ray probe pulse with a duration of 100~ps was directed at the sample in a near collinear arrangement with the laser pump pulse.  X-ray diffraction was observed in a symmetric geometry (a Bragg-Brentano or $\theta - 2 \theta$ configuration in which the diffraction vector is always normal to the sample surface) and the (111) and (222) reflections were probed as a function of fluence. The comparatively longer duration of the probe pulse relative to the pump pulse resulted in a corresponding convolution in time in the measured signal.  The Ti-sapphire amplified laser system produced pulses with energies up to 5~mJ/pulse synchronous to the ring; the jitter of the laser with respect to the x-ray pulses was less than 5~ps. By varying the time difference $\delta t$ between the pump and probe pulses, the structural dynamics of the film as a function of time could be observed; the $t=0$ was defined as the maximum temporal overlap of the pump and probe pulses using a high-speed photodiode. The dynamics were measured in 10 ps steps on both sides of $t=0$ for $\delta t$ up to 60 ps and with increasing step sizes for $\delta t$ up to several nanoseconds. The experiment was repeated for a range of pump fluences  \mr{8, 10,12, 16 \, and \, 24 \, mJ/cm^{2}}  and for a variety of $\delta t$ after the determination of the amorphization threshold of $\mathrm{54 \, mJ/cm^{2}}$.  A collection of representative snapshots of the time evolution of the (222) diffraction peak is shown in Fig. \ref{fig1}. 

\subsection{Experimental Results}

\begin{figure}[h] %  figure placement: here, top, bottom, or page
   \centering
   \includegraphics[width=8 cm]{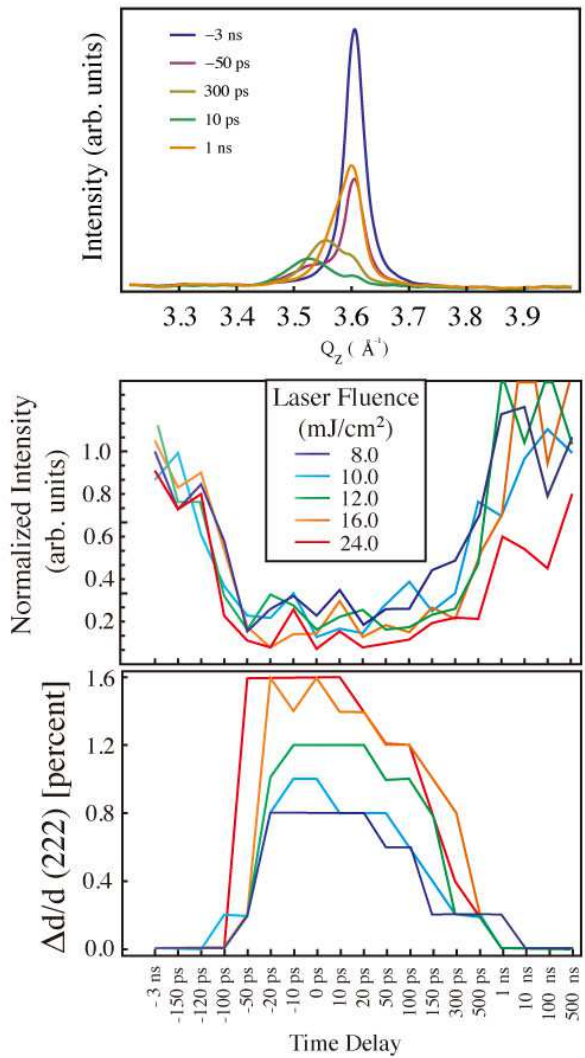} 
   \caption{(Color Online) Upper Panel: A plot of the intensity distribution for a few prototypical values of $\delta t$ for \mr{16 \, mJ/cm^{2}}. Middle Panel: A plot of the integrated intensity of the GST (222) peak as a function of $\delta t$ for different fluences.  Lower Panel: A plot of the fractional shift in the (222) plane spacing in percent as a function of $\delta t$ for different fluences. To obtain the pseudo 1D spectra for all $\delta t$ values and fluences, the diffraction peaks were fit by Gaussian functions and the results tabulated.}
\label{fig2}
\end{figure}

For all fluences, a similar trend was observed.  A decrease in the (222) peak intensity was seen for $\delta t$ larger than \mr{\sim -100 \,ps} with a maximum decrease in intensity of 80-90\% at about 10 ps.   At that time about 10\% of the initial intensity is redirected into a sideband peak at lower $Q_{z}$. The (222) peak subsequently recovered to its initial value after 1 ns as can be seen in Fig. \ref{fig2}.  A nearly identical time dependence for the fractional change in d-spacing $\Delta d/d$ along the (111) direction was found with the magnitude of the shift growing larger with increasing laser fluence although the maximum decrease in intensity was less than 30\%.
For the (222) peak, the maximum value for $\Delta d/d$ of 1.6\% was measured for the highest laser fluence of \mr{24 \, mJ/cm^{2}}.  

We mention here in passing that the reduction in diffraction peak intensities due to the experimentally determined Debye-Waller factor of GST accounts for only $\mathrm{\sim 20\%}$ reduction in peak intensity even at the melting point and cannot account for the effects reported here.\cite{Matsunaga09}  It should also be mentioned here that the time-resolved behavior of the (111) reflection was similar to that observed for the (222) reflection, however, due to the limited angular dispersion of the (111) peak, the intensity reduction was markedly less and no discernible shift could be resolved. We choose to focus on the (222) reflection in this paper.

\begin{figure}[h!] %  figure placement: here, top, bottom, or page
   \centering
   \includegraphics[width=8.8cm]{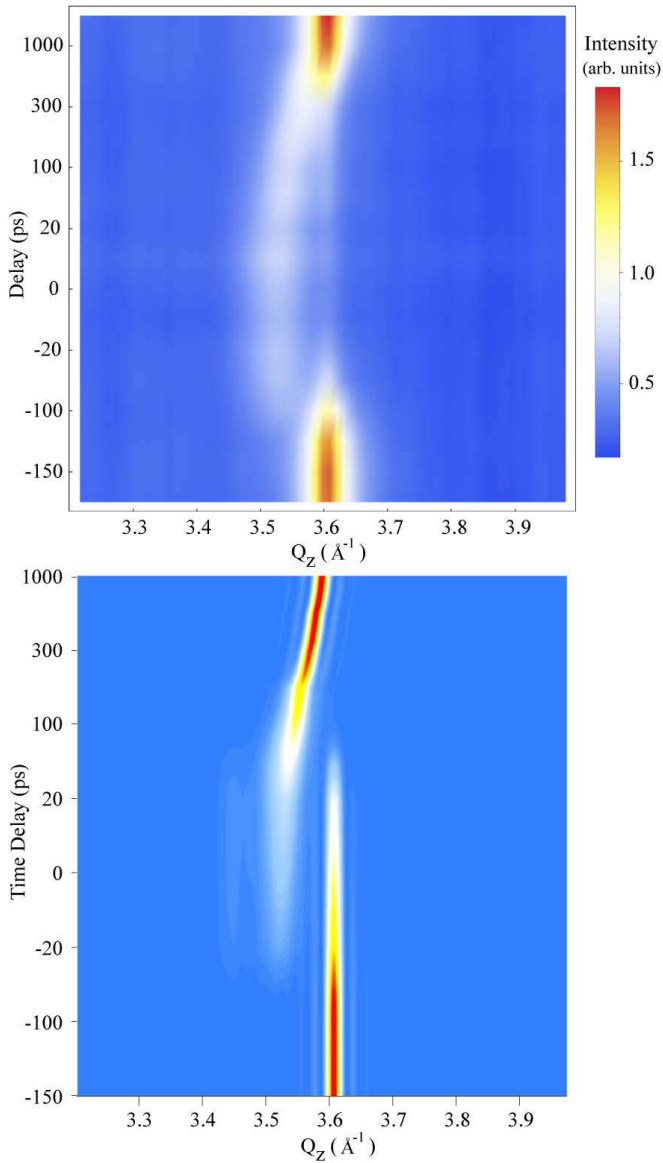} 
   \caption{(Color Online) Upper Panel: Measured time-$Q_{z}$ intensity plot of the GST (222) diffraction peak for a fluence of $\mathrm{16 \, mJ/cm^{2}}$. Lower Panel: Simulated time-\mr{Q_{z}} intensity plot of the GST (222) diffraction peak carried out using a ball and spring model (see text for details). The \mr{Q_{z}} axis in the figure corresponds to the reciprocal space direction $\left <111 \right >^{*}$ while the \mr{Q_{x}} axis corresponds to the reciprocal space $\left <110 \right >^{*}$ direction.}
   \label{fig3}
\end{figure}

Fig.~\ref{fig3} (upper panel) shows the intensity of the GST (222) reflection as a function of time for a pump fluence of \mr{16 \, mJ/cm^{2}}.  For times before the laser pulse, the Laue diffraction condition is given by \mr{\mathrm{\overrightarrow{Q}=\overrightarrow{G}}}, where $\overrightarrow{Q}$ represents the momentum transfer of the elastic scattering process and \mr{\overrightarrow{G}} is a reciprocal lattice vector.  The pump-pulse induces a fast displacement of the equilibrium atom positions in the GST film.  Providing the duration of the pump-pulse is significantly shorter than the phonon period, phonons in phase (coherent) can be generated.\cite{Zeiger92}  The change in equilibrium positions gives rise to a distribution of acoustic CP whose wavelength is centered at twice the film thickness due to the boundary conditions imposed by the much lower optical absorption of the silicon substrate and $\mathrm{Si_{3}N_{4}}$ cap/film interface. The stress leading to this fast displacement can be written in terms of the contributions of the electronic and lattice Gr{\"u}neisen parameters or $\sigma \sim \gamma_{e} C_{e} T_{e} + \gamma_{L} C_{L} T_{L}$ where $\gamma_{i}, C_{i}, \, \mathrm{and} \, T_{i}$ represent the Gr{\"u}neisen parameter, the specific heat, and the temperature for the electronic ($i=e$) and lattice ($i=L$) subsystems. The dimensionless Gr{\"u}neisen parameter describes the change in length or volume required to minimize the free energy of the system in response a temperature change.  Gradients in the stress $\sigma$ that occur at the interfaces of the GST layer give rise to propagating elastic waves.\cite{Park05,Thomsen86} The electronic stress given by $\gamma_{e}$ plays a significant role in the early stage of thermal expansion before the electrons and lattice reach thermal equilibrium, which is typically less than a few picosecond later.

 The presence of a longitudinal coherent phonon with wavevector \mr{\overrightarrow{q}} modifies the diffraction condition so that the quasi-elastic Laue condition becomes $\mathrm{\overrightarrow{G}=\overrightarrow{Q}+\overrightarrow{q}}$ where $\overrightarrow{q}$ represents the effective wavevector of the CP.  This gives rise to a transient displacement of the GST (222) reflection while the propagating strain field due to the CP is present in the GST epilayer.  

\begin{figure}[h!] %  figure placement: here, top, bottom, or page
   \centering
   \includegraphics[width= 8.8cm]{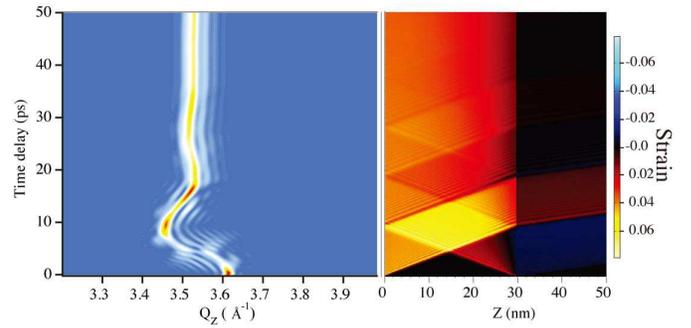} 
   \caption{(Color Online) Left Panel:  Simulated time-$Q_{z}$ intensity plot of the GST (222) diffraction peak carried out using a ball and spring model (see text for details) in the vicinity of $t=0$ \emph{without} convolution effects. The intensity scale is the same as in Fig. 2. Right Panel: A plot of the one-dimensional strain in percent as a function of time as calculated by a ball and spring model.}
   \label{fig4}
\end{figure}

Fig.~\ref{fig3} (lower panel) shows the simulated time dependence of the GST (222) diffraction signal calculated using the one-dimensional linear chain model \texttt{udkm1Dsim}. \cite{Schick14}  The model has yielded excellent fits of ultrafast phonon dynamics in metal/insulator superlattice systems.\cite{Herzog12} For the purpose of comparison, the model spectra plot consists of a stack of simulated spectra for the times at which the experimental data was taken, namely  -150~ps, -120~ps, -100~ps, -50~ps, -20~ps, -10~ps, 0~ps, 10~ps, 20~ps, 50~ps, 100~ps, 150~ps, 300~ps, 500~ps, 1~ns, and 10~ns. In the figure, the maximum time duration has been limited to 1 ns. In addition, the simulated spectra have been convoluted along the time axis by a 100 ps Gaussian kernel to account for the finite duration of the x-ray pulse.  In the model, each GST unit cell is conceptualized as a ball with identical springs connecting each ball in a one-dimensional chain; the mass of each ball is given by the sum of the masses of atoms in the unit cell which was assumed to have the rock salt structure.\cite{Yamada00} The substrate was modeled in a similar manner with appropriate values for the  ball mass and spring force constants.  The springs force constant is given by $k_{i}=m_{i} v_{i}^{2}/a_{i}^2$ where $v_{i}$ is the longitudinal sound velocity, $m_{i}$ is the mass of a unit cell, and $a_{i}$ is the out-of-plane lattice constant of the \emph{i}th unit cell.  In \texttt{udkm1Dsim}, it is assumed that at time $t=0$, the laser pulse induces instantaneous stress due to a change in the equilibrium position of atoms to a wider spacing as determined by the coefficient of thermal expansion and the temperature rise at that point induced by the absorbed light. The stress field is assumed to be applied instantaneously at $t=0$ and the equations of motion are solved to predict the subsequent evolution of the system. While in the harmonic approximation, the equations of motion can be diagonalized to give the normal modes of the system, the inclusion of heat diffusion effects requires numerical solution of the equations of motion as was done here. The time period required for the GST (222) reflection to revert to its unperturbed position was on the order of several nanoseconds implying that the effects of heat diffusion for time scales less than 100 picoseconds when the most intense dynamics occur were negligible.
  The initial stress relative to the arrival of the laser pulse is assumed to be proportional to the temperature increase at a given depth which is, in turn, calculated taking into account the temperature-dependent specific heat of the material and the (depth dependent) heat due to the absorbed light as given by the Lambert-Beer law for each material.   The various parameters used as input to \texttt{udkm1Dsim} are summarized in Table~\ref{tab:params}.  Experimentally determined temperature-dependent values of thermal conductivity $\kappa$, and heat capacity $c_{p}$ were used for GST and hence are omitted from the table.\cite{Kuwahara07} For each time step $\delta t$, the Takagi-Taupin dynamical x-ray scattering equations are solved to determine the change in diffracted intensity. Finally, as mentioned above, the simulated results are convoluted in time with a Gaussian kernel to take into account the $\mathrm{\sim 100 \, ps}$ duration of the x-ray pulses of the 16-bunch mode used.  Fig. \ref{fig3}, lower panel shows the simulated x-ray diffraction pattern for the GST (222) reflection as a function of time calculated using the procedure described above. 

\begin{table}[htdp]
\caption{Parameters used as input in the udk1Dsim simulation.  Here $a$, $v_{s}$, $\alpha_{L}$, and $\alpha$ represent the lattice constant, the sound velocity,  the linear expansion constant, and the optical penetration depth at $\mathrm{\lambda=800\, nm}$, respectively.  The references for Silicon are lattice constant \cite{Palik85}, sound velocity \cite{Kim93}, linear expansion coefficient \cite{Swenson83}, and optical penetration depth \cite{Palik85}. The references for GST are lattice constant \cite{Matsunaga04} (as also verified by the unexcited diffraction data from this experiment) and sound velocity \cite{Schick11}. See the text for an explanation for the choices of linear expansion coefficient and optical penetration depth.}
\begin{ruledtabular}
\begin{center}
\begin{footnotesize}
\begin{tabular}{ccccccc}Material & $a$ & $v_{s}$ & $\alpha_{L}$ & $\mathrm{\alpha}$ \\ 
& (\AA) & (nm/ps) & $\mathrm{(K^{-1})}$ & ($\mathrm{cm}$) \\
Silicon & 5.431& 9.35 & $\mathrm{2.6 \times 10^{-6}}$ & $\mathrm{1.07 \times 10^{-3}}$ \\
GST &  6.0293 & 3.19 &  $\mathrm{{7.2 \times 10^{-5} }}$ & ${1.05 \times 10^{-5}}$\end{tabular}
\end{footnotesize}
\end{center}
\label{tab:params}
\end{ruledtabular}
\end{table}%

 It bears repeating that although the simulation was carried out with a very fine time step of 1 ps, the result as shown in Fig.~\ref{fig3} (lower panel) does not show diffracted intensity over a continuous time domain, but rather consists of stacks of snapshots of the simulated results at the experimentally observed (discrete) time intervals. Fig. \ref{fig4} (left panel) shows the simulation (without convolution) and a picosecond time step in the vicinity of $t=0$ and Fig.~\ref{fig4} (right panel) illustrates the calculated strain field as a function of time and depth into the sample.  It should be noted here that the strain field is not uniform and the shift in the x-ray diffraction peak represents an average-like displacement obtained by numerical solution of the x-ray dynamical diffraction equations.
  As can be seen in the strain figure, differences in absorption at the cap-sample and sample-substrate boundaries generate large gradients in the relative displacement at the interfaces and lead to the generation of compressive strain waves originating from both interfaces that travel into the bulk of the sample. The maximum strain is reached for a time $t_{m}$ given by $d/v_{s}$ where $d$ is the film thickness and $v_{s}$ is the longitudinal sound velocity.  The maximum displacement corresponds to the transit time of the compression wave from each interface to the center of the film.  For the case of GST, this is given by the film thickness divided by the sound velocity or $\mathrm{30\, nm/(3.19\, nm/ps) = 9.4 \,ps}$ providing excellent agreement with experiment. 
  
\section{Discussion}
While the simulation overall is in relatively good agreement with experiment, discrepancies with room temperature parameter values of absorption length and thermal expansion coefficient are present, that in turn suggest the presence of excited state effects in addition to thermal effects as has been suggested in the literature. \cite{Li11} In general in the \texttt{udkm1Dsim} model, the magnitude of the gradient of the thermal expansion at the interfaces leads to the generation of a expansion wave in GST associated with an acoustic phonon.  This implies that the magnitude of the shift in the x-ray signal along $Q_{z}$ is proportional to the induced temperature rise due to the absorbed light with higher laser fluences leading to larger shifts.  As GST is a well known phase-change alloy with a melting point of approximately 900~K, a transient rise of temperature significantly above the melting point, would be expected to lead to the formation of an amorphous phase due to the excellent thermal contact of the epilayer to the silicon substrate. This conclusion that there is no melting present is further supported by the lack of a concomitant drop in the intensity of the (111) peak.  As the diffraction peak is experimentally observed to return to its original intensity after a period on the order of a nanosecond, the formation of an amorphous phase can be excluded leading to the conclusion that the maximum temperature rise must be less than the melting point of GST.  To satisfy this condition, we thus selected an optical penetration depth value that limited the temperature rise to a few degrees below the melting point and increased the thermal expansion coefficient to match the experimentally observed displacement along $Q_{z}$. Calculations assuming the experimental value of optical absorption resulted in nearly all of the incident flux being absorbed by the GST film leading to a temperature rise of over 3600~K, a temperature that is completely inconsistent with the diffraction peak returning to its initial position within a nanosecond.  An effective fluence value of \mr{8 \, mJ/cm^{2}} was used as the experimentally measured reflectivity at 800~nm is approximately 0.5.  A calculation of the reflectivity using the Fresnel equations and experimental values of refractive index and angle of incidence yields a reflectivity value consistent with this observation of 49\%.  This approach led to the values of the linear expansion parameter $\alpha_{L}$ and optical penetration length $\alpha$ shown in Table \ref{tab:params}. The value used for the optical penetration depth that maintained the maximum temperature constraint is a factor of approximately six times larger than the published value of  \mr{\alpha \sim 1.9\times 10^{-6}\, cm}.~\cite{Lee08} This large difference from the literature value may stem from either dynamic changes in absorption due to the short duration of the pump pulse, e.g. photo-bleaching or from alternative energy loss mechanisms such as energy transport into the substrate via ballistic electrons.  Photo-bleaching or a transient reduction in optical absorption is known to occur for short intense pump pulses due to several factors including bandgap renormalization due to the accumulation of carriers at the band edges and transient increases in reflectivity due to the creation of an electron-hole plasma driven by the electrical field of the optical pump pulse. \cite{Kudryashov07,Srivastava04} 

We have measured the transmission (not shown) of 800 nm, 40 fs pulses through a 20 nm thick polycrystalline GST layer.  The absorption constant calculated using the literature value of the reflectivity shows two linear regimes, the first from zero to about \mr{8 \, mJ/cm^{2}} and a second from approximately \mr{8 \, mJ/cm^{2}} to about \mr{30 \, mJ/cm^{2}}. The slope of the absorption versus input fluence decreases by roughly a factor of three between the two regimes. Note that the peak power in the current experiment ($\mathrm{114 \, TW/m^{2}}$) is below that corresponding to the \mr{8 \, mJ/cm^{2}} threshold ($\mathrm{2 \, PW/m^{2}}$) reported above taking into account the ratio of pulse durations or $\mathrm{700\, fs/40\, fs \sim 17}$.  No quadratic dependence on fluence was found suggesting that two photon absorption did not play a significant role in the absorption process in the current experiment.  The underlying mechanism for the non-linear absorption effects in GST has been suggested to be complex \cite{Liu11} and is beyond the scope of the current work.  The effects of non-linear absorption have been accounted for by means of fitting the optical penetration depth.
 
The lattice expansion that in turn triggers the creation of acoustic phonons can be considered to have two origins as described by the Gr{\"u}neisen parameter.   Both the lattice and excited carriers contribute to lattice expansion.  For ultrafast processes such as in the current experiment, the electron and lattice subsystems are not initially in equilibrium leading to two independent contributions (and two different subsystem temperatures) to lattice expansion as exemplified by the electronic and lattice Gr{\"u}neisen parameters, respectively.\cite{Wang13}  
The more than factor of four increase in the thermal expansion coefficient $\alpha_{L}$ over the published value~\cite{Park08} of \mr{1.74\times 10^{-5} \, K^{-1}} found in the current fitting may reflect the contributions of the  electronic Gr{\"u}neisen parameter to the induced stress in the film.   As the system can only expand along the z direction, an increase of up to a factor of three in $\alpha_{L}$ might be anticipated due to the decrease in dimensionality but, the factor of four found here suggests additional contributions from electronic excitation effects.\cite{Kolobov11}  Although the electron system driven lattice expansion may be transient until the electrons recombine or escape the layer into the substrate, the electronic Gr{\"u}neisen parameter can still serve as a significant source of strain in the generation of an acoustic wave. 
While there is no publication that shows the thermal expansion of GST over a wide temperature range due to the presence of an irreversible phase transformation at about 450 K, there is data published on a closely related alloy $\mathrm{Ge_{8}Sb_{2}Te_{11}}$ which shows a nearly linear expansion from 80 K to just below the melting point.\cite{Matsunaga08}  This strongly suggests that the lattice G{\"u}neisen parameter (the relationship between lattice thermal expansion and the lattice specific heat) for GST is linear and that the need to increase the linear expansion factor in the simulation is a consequence of the nonzero value of the electronic Gr{\"u}neisen parameter.

\section{Conclusions}

In conclusion we have measured and successfully modeled the atomic displacement of atoms in GST in the presence of coherent acoustic phonons. A ball and spring model was used to simulate the displacement field as a function of time and the corresponding shifts in the (222) x-ray diffraction peak. In order to model the experimental data under the constraint that the GST film does not melt, it was necessary to modify both the optical penetration depth and thermal expansion coefficients from their literature values.  The increase in the absorption length parameter suggests the presence of either alternative energy loss mechanisms such as ballistic electron transport or photo bleaching effects due to the ultrashort pulses used.  As the absorption length parameter was thus constrained, the increase in the thermal expansion factor can be considered to be nominally uncorrelated with the absorption length parameter in the fitting process.  The increase in the thermal expansion factor strongly suggests the presence of electronic Gr{\"u}neisen effects due to transient electron heating that are comparable in magnitude to the lattice-temperature induced expansion.  The presence of coherent acoustical phonons and their successful observation are a first step in inducing local displacements in a controlled manner that may lead to non-thermal transitions between bonding states in GST.   While further studies with better time resolution are necessary to fully understand the role of electronic excitation, a clear contribution of electronic Gr{\"u}neisen effects was observed. To attain the goal of using coherent phonons to drive GST between bonding states, shorter wavelength optical phonons are necessary and these studies are underway.

\begin{acknowledgments}
We would like to acknowledge approval of an ESRF proposal SI-2346 at beamline ID9 and the help of Michael Wulff and Dmitry Khakhulin. P.F., K.M. , and M.H. (tsukuba) would also like to acknowledge support from the X-ray Free Electron Laser Priority Strategy Program projects 12013011 and 12013023 from the Ministry of Education, Science, Sports, and Culture of Japan. We are grateful to S. Behnke and C. Stemmler for their dedicated maintenance of the MBE system and to O. Bierwagen for support.  We would also like to acknowledge numerous productive email exchanges with Daniel Schick in relation to \texttt{udkm1Dsim}.

\end{acknowledgments}

%\bibliography{bibliography}

\end{document}